\documentclass[prb,twocolumn,showpacs]{revtex4}

\usepackage{amsmath,amsfonts,amssymb,bm}
\usepackage{dcolumn}
\usepackage[final]{graphicx}
\usepackage{bm}

\begin{document}

\bibliographystyle{apsrev}

\title{Optimal quantum gates for semiconductor qubits}

\author{Ulrich Hohenester}\email{ulrich.hohenester@uni-graz.at}
\affiliation{Institut f\"ur Physik,
  Karl--Franzens--Universit\"at Graz, Universit\"atsplatz 5,
  8010 Graz, Austria}

\date{Octotber 11, 2006}

\begin{abstract}

We employ optimal control theory to design optimized quantum gates for solid-state qubits subject to decoherence. At the example of a gate-controlled semiconductor quantum dot molecule we demonstrate that decoherence due to phonon couplings can be strongly suppressed. Our results suggest a much broader class of quantum control strategies in solids.

\end{abstract}



\maketitle


Decoherence is the process within which a quantum system becomes entangled with its environment and loses its quantum properties. It is responsible for the emergence of classicality~\cite{zurek:03} and constitutes the main obstacle in the implementation of quantum computers.~\cite{nielsen:00,bennett:00} For atoms, environment couplings can be strongly suppressed by working at ultrahigh vacuum and ultralow temperature. For artificial atoms ---the solid-state analogies to atoms---, things are more cumbersome because they are intimately incorporated in the surrounding solid-state environment and suffer from various decoherence channels. A number of quantum control techniques are known, such as quantum bang-bang control,~\cite{viola:98} decoherence-free subspaces,~\cite{zanardi:97} or spin-echo pulses,~\cite{petta:05} that allow to fight decoherence. However, it is not the system--environment interaction itself that leads to decoherence, but the imprint of the quantum state into the environmental degrees of freedom: the environment measures the quantum system. In this paper we show that optimal control theory~\cite{peirce:88,rabitz:00} allows to design control strategies where quantum systems can be controlled even in presence of such environment couplings without suffering significant decoherence. This opens the possibility for a much broader class of quantum control that might render possible high-performance quantum computation in solids.

An attractive candidate for a solid-state qubit is based on semiconductor quantum dots, which allow controlled coupling of one or more electrons by means of voltage pulses applied to electrostatic gates.~\cite{wiel:03} The spin of electrons confined in such dots provides a viable quantum memory owing to its long life and coherence times of the order of micro to milliseconds.~\cite{kroutvar:04,petta:05} In the seminal work of Loss and DiVincenzo~\cite{loss:98} a mixed quantum computation approach had been envisioned, where the quantum information is encoded in the spin degrees of freedom, thus benefiting from the long spin coherence times, and the much stronger coupling to the charge degrees of freedom is exploited for performing fast quantum gates. Recent experiments have indeed demonstrated the coherent manipulation of charge states in coupled dots.~\cite{hayashi:03,petta:04,petta:05} Electron charge, however, not only couples to the external control gates but also to the solid-state environment, e.g. to phonons, which introduces decoherence during gate manipulations. Theoretical work has estimated for realistic quantum dot structures that typically ten to hundred quantum gates can be performed within the charge coherence time,~\cite{vorojtsov:05,stavrou:05} which provides a serious bottleneck for solid state based quantum computation. In addition, future miniaturization of nanostructures will result in a further increase of such decoherence owing to the larger number of phonon modes to which carriers in small dots can couple.~\cite{krummheuer:02}

\begin{figure}[b]
\centerline{\includegraphics[width=0.9\columnwidth]{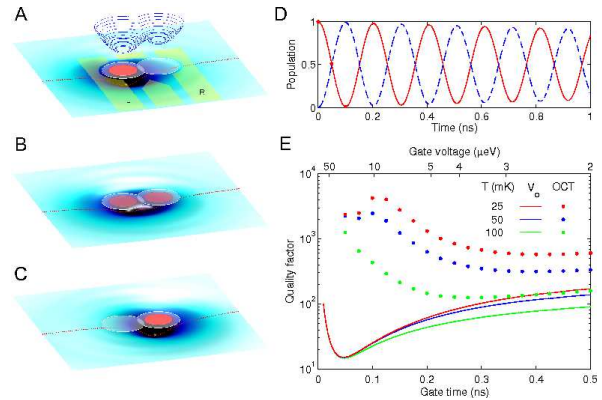}}
\caption{
(a) Schematic sketch of the double dot structure considered in our calculations. By applying voltage pulses to the top gates $L$, $T$, and $R$ the tunnel coupling and energy detuning can be controlled. We consider dot radii of 60 nm, a dot distance of 80 nm, and GaAs-based material parameters \cite{vorojtsov:05}. The surface plot at the bottom of panel (a) indicates the lattice distortation for the electron in the left dot. Upon applying a tunnel coupling $v_0$ the electron tunnels from the left to the right dot. Panels (b) and (c), respectively, show snapshots where the electron is delocalized over the whole structure and localized in the right dot. (d) Population of the left (solid line) and right (dashed line) dot state. The oscillation is damped because of phonon-assisted dephasing. (e) Quality factor as a function of gate time for three different temperatures and for constant pulses $v_0$ (lines) and optimized pulses (symbols).
}
\end{figure}

Let us consider the setup depicted in fig.~1a where a single electron is confined in a double dot structure. Although true quantum algorithms will involve the manipulation of two or more electrons, the case study of a single electron completely suffices to understand the problems inherent to charge control within the solid state. The tunnel coupling $v(t)$ and the energy detuning $\varepsilon(t)$ between the left and right dot can be controlled through voltage pulses applied to the top gates $L$, $T$, and $R$, and the hamiltonian describing the system is of the form
\begin{equation}\label{eq:ham-twolevel}
  H_0=\left(\begin{array}{cc} 0 & v(t) \\ v(t) & \varepsilon(t) \\
  \end{array}\right)\,.
\end{equation}
Suppose that the electron is initially in the left dot. When at time zero a constant tunnel coupling $v_0$ is turned on, the left-dot and right-dot states become coupled and the electron will start to tunnel back and forth between the two dots, as experimentally demonstrated.~\cite{hayashi:03} However, electrons in solids always interact with the lattice degrees of freedom, which, for the confined electron states under consideration, results in a slight deformation of the lattice in the vicinity of the electron. Such interaction is conveniently described within the independent Boson model~\cite{mahan:81,brandes:05}
\begin{equation}\label{eq:ham-indboson}
  H_{\rm ph}=\sum_{\bm q}g_q
  \left(b_{\bm q}^\dagger+b_{-\bm q}^{\phantom\dagger}\right)
  \,\left(\begin{array}{cc} s_{\bm q}^L & 0 \\ 0 & s_{\bm q}^R \\
  \end{array}\right)\,,
\end{equation}
where $\bm q$ is the wavevector of the quantized eigenmodes of the lattice, i.e. phonons, $g_q$ the coupling constant of the bulk material (piezoelectric and deformation potential), $b_{\bm q}^\dagger$ the bosonic creation operator for phonons, and $s_{\bm q}^i$ the usual form factor for the electron-phonon coupling in the left or right dot.~\cite{vorojtsov:05,brandes:05} Through the different form factors $s_{\bm q}^i$, the electron couples differently to the phonons in the left and right dot, respectively. The surface plot at the bottom of fig.~1a shows the lattice displacement~\cite{mahan:81}
\begin{equation}\label{eq:displacement}
  u(\bm r)=\sum_{\bm q}\left(2\rho\omega_q\right)^{-\frac 12}\,e^{i\bm q\bm r}
  \langle b_{\bm q}\rangle+\mbox{c.c.}\,
\end{equation}
for the electron initially localized in the left dot as computed within a density-matrix framework,~\cite{rossi:02,foerstner:03a,hohenester.prl:04,hohenester.review:06,remark.densitymatrix} with $\rho$ the mass density of the semiconductor, $\omega_q=cq$ the phonon energy, and $c$ the sound velocity. Conversely, when the electron is localized in the right dot, fig.~1c, the lattice becomes distorted around the right dot. In a sense, this finding is reminiscent of molecular physics where electronic excitations are accompanied by variations of the molecular structure, though in our case the coupling is much weaker and to a continuum of phonon modes rather than to a few vibronic states.

\begin{figure}
\centerline{\includegraphics[width=0.75\columnwidth]{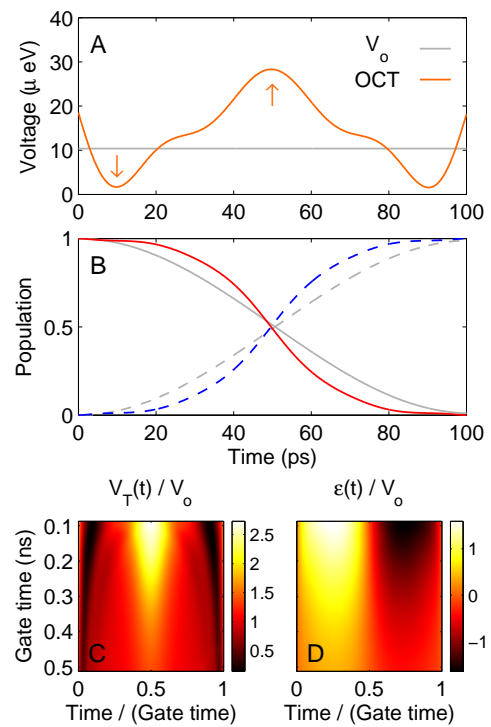}}
\caption{
Results of optimal control calculations for a lattice temperature of 50 mK. (a) Optimized voltage pulse $v(t)$ and (b) time evolution of left-dot (solid line) and right-dot (dashed line) population, for a gate time of 100 ps. (c) Density plot of the optimized pulses $v(t)$, in units of the constant pulse height $v_0$, for different gate times. (d) Same as (c), but for $\varepsilon(t)$.
}
\end{figure}

\begin{figure*}
\centerline{\includegraphics[width=1.5\columnwidth]{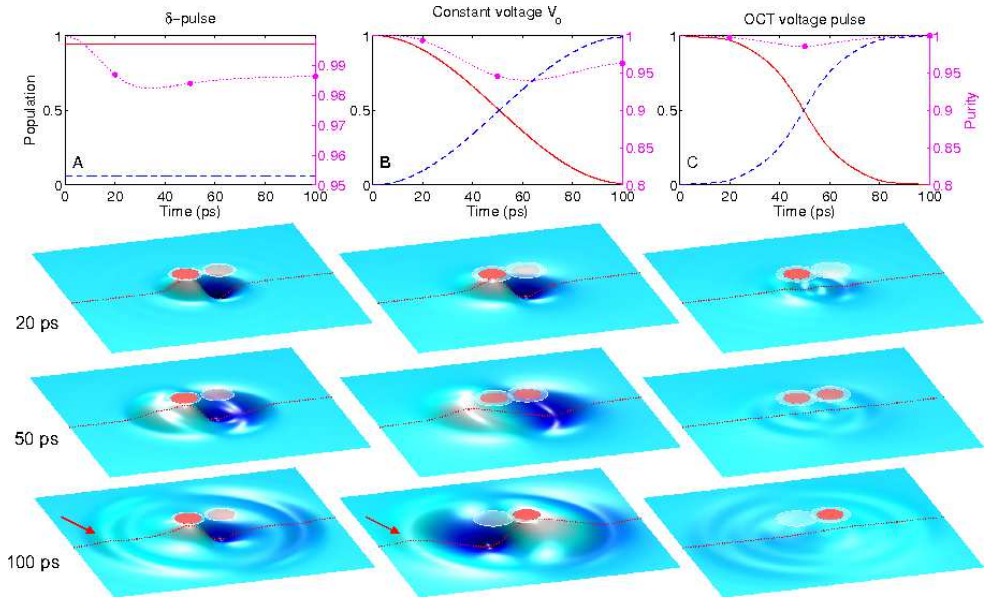}}
\caption{
Time evolution of left-dot (solid lines) and right-dot (dasehed lines) population, and of purity (dotted lines), for (a) $\delta$-like excitation at time zero, (b) constant voltage pulse $v_0$, and (c) optimized control fields $v(t)$ and $\varepsilon(t)$. The lower panels show real-space maps of the electron-phonon entanglement (for definition see text) at three selected times.
}
\end{figure*}

As we will show next, this coupling to a phonon continuum has a drastic influence on the coherent charge oscillations. Upon turning on the tunnel coupling $v_0$ between the two dots, the electron starts to oscillate and the lattice distortion follows, as shown in figs.~1a--c. However, since the phonon cloud cannot follow instantaneously due to the finite phonon frequencies $\omega_q$, part of the quantum coherence is transferred from the electron system to the phonons, resulting in a coherence loss of the electron motion as evidenced by the damping of the oscillations shown in panel (d). To quantify such loss, we introduce in accordance to ref.~\onlinecite{vorojtsov:05} the quality factor $Q$ that determines how many charge oscillations can be resolved within the decoherence time. We shall also find it convenient to refer to the electron transport from the left to the right dot as a {\em quantum gate},\/ with the gate fidelity~\cite{nielsen:00} being directly given by $Q$. Figure 1e reports the dependence of $Q$ on gate time and lattice temperature.~\cite{remark.densitymatrix} Over a range of experimentally accessible temperatures $Q$ has a minimum for gate times around 50 ps, and is significantly enhanced at both shorter and longer times. The appearance of this minimum has been discussed in length in ref.~\onlinecite{vorojtsov:05} and can be qualitatively understood as follows. At short gate times $T$ the electron moves on a timescale much shorter than the response time of the phonon cloud, and consequently the electron transport is not affected by the much slower lattice dynamics (dynamic decoupling): here $Q$ increases with decreasing $T$. On the other hand, at long gate times the phonon cloud follows almost adiabatically, and $Q$ increases with increasing $T$. The minimum occurs at a time $\tau\sim\omega_q^{-1}$ where the phonon wavevector $q\sim 2\pi/d$ matches the interdot distance $d$, as will be discussed in more detail below. This behavior of $Q$ suggests that coherent electron transport should be much faster or slower than $\tau$, which imposes serious constraints on quantum gates. In reality the situation is even more adverse. For fast gating other quantum dot states might become excited, wheras for slow gating additional environment couplings might gain importance. Also a further miniaturization of the double-dot structure will lead to a further decrease of $Q$.

In ref.~\onlinecite{hohenester.prl:04} we showed for the optical control of quantum dot exciatations that such phonon-assisted decoherence can be strongly suppressed through laser-pulse shaping. Accordingly, we might expect that for the tunnel-coupled double dot an optimization of the control fields $v(t)$ and $\varepsilon(t)$ could improve the quantum-gate performance. In the following we thus employ the framework of optimal control theory~\cite{peirce:88,rabitz:00} to search for control fields $v(t)$ and $\varepsilon(t)$ that maximize $Q$, i.e., we are seeking for voltage pulses that minimize decoherence losses during gating. Optimal control theory (OCT) accomplishes the search for optimized control fields by converting the constrained minimization to an unconstrained one, by means of Lagrange multipliers, and formulating a numerical algorithm which, starting from an initial guess for the control fields, succeedingly improves them. Details of our numerical approach can be found in refs.~\onlinecite{borzi.pra:02,hohenester.prl:04,hohenester.review:06}. Figure 2 shows results of our OCT calculations. The optimized $v(t)$ differs from the constant $v_0$ one in that the strength is strongly reduced at the beginning and at the end of the quantum gate, and enhanced in the middle (indicated by arrows). At the same time, the energy offset $\varepsilon(t)$ [see panel (d)] is varied during the gate from positive to negative detuning. Panels (c) and (d) report that these control strategies prevail over a wide range of gate times. As apparent from fig.~2b, the transfer process from the left to the right dot is not drastically altered by the optimized fields $v(t)$ and $\varepsilon(t)$ in comparison to that of the constant field $v_0$ (gray lines). On the other hand, the quality factor $Q$ of the OCT gates [symbols in fig.~1(e)] becomes boosted by several orders of magnitude when using optimized pulses. Even more striking is that the OCT fields perform best for gate times where the constant $v_0$ field performs worst. {\em Thus, optimal control theory allows to design control strategies that can drastically outperform more simple schemes.}

To understand the drastic improvement of optimized pulses, in fig.~3a we first analyze the more simplified situation of a $\delta$-like voltage pulse where at time zero the electron is brought instantaneously from the left dot to a superposition state between the two dots, and the dot coupling is turned off at later times. Consequently, the left- and right-dot populations shown in fig.~3a remain constant. We additionally plot the purity ${tr}\bm\rho^2$ of the electron system, which, starting from the initial value of one, gradually decreases, thus indicating the transition from an inital pure state to a final mixture. Similar to the entropy, the purity is a measure of the degree of entanglement between the electronic and phononic system, i.e., how much information the phonons posess about the quantum properties of the electron state. In analogy to eq.~\eqref{eq:displacement} we define
\begin{equation}\label{eq:entanglement}
  u_1(\bm r)=\sum_{\bm q}\left(2\rho\omega_q\right)^{-\frac 12}\,e^{i\bm q\bm r}
  \langle\!\langle b_{\bm q}\sigma_1\rangle\!\rangle+\mbox{c.c.}\,
\end{equation}
as an entanglement measure, with $\langle\!\langle b_{\bm q}\sigma_1\rangle\!\rangle=\langle b_{\bm q}\sigma_1\rangle-\langle b_{\bm q}\rangle\langle\sigma_1\rangle$ the correlation between phonon mode $b_{\bm q}$ and the (real part) quantum coherence $\sigma_1$.~\cite{remark.densitymatrix} When at time zero the electron superposition state is apruptly prepared, it requires a time $\tau\sim\omega_q^{-1}$ for the phonons to aquire information about the modified electron state. In particular phonon modes with wavevector $2\pi/d$, where $d$ is the interdot distance, carry information about the superposition properties of the electron state, and thus set the timescale for the entanglement buildup. The snapshots of $u_1(\bm r)$ in the left column of fig.~3 report such buildup in the vicinty of the dots. However, due to the phonon inertia the lattice distortion overshoots, instead of smoothly approaching the new equilibrium position, and a phonon wavepacket is emitted from the dots (see arrow) \cite{brandes:05,truegler:05} which imprints the quantum information about the superposition state into the environment and thus reduces its quantum properties: the system sufferes decoherence. Similar behaviour of overshooting and wavepacket emission is observed for the quantum gate with constant $v_0$. In contrast, the optimized quantum gate shown in the right column of fig.~3 strongly suppresses the emission of a phonon wavepacket by reducing the tunnel coupling between the dots in the initial stage of the electron transfer, see arrow in fig.~2a, and thus allows the lattice to react smoothly to the time varying electron configuration. Similar conclusions hold for the final stage of the transfer. As regarding the peak in the middle of the control pulse shown in fig.~2a, we find that its shape strongly depends on the detailed properties of the phonon coupling $g_q$, and is strong in case of piezoelectric coupling and absent in case of deformation-potential coupling. 

Our optimal quantum control strategy differs appreciably from other control strategies. The inherent coupling of electrons to phonons excludes quantum state manipulations in decoherence free subspaces~\cite{zanardi:97} or other quantum-optical control techniques, such as, e.g., stimulated Raman adiabatic passage,~\cite{bergmann:98} where quantum state transfer is achieved through states fully decoupled from the environment. Furthermore, the time dynamics of the phonon degrees of freedom disables spin-echo techniques to restore pure quantum states by means of effective time reversal through $\pi$ pulses. Finally, optimal control only requires smooth voltage variations on the timescale of tens of picoseconds, rather than sub-picosecond pulses needed for quantum bang-bang control,~\cite{viola:98} where the system has to become dynamically decoupled from the environment. Such short pulses, which are at the frontier of presentday technology, have been also proposed for other quantum control applications,~\cite{calderon:06} and might introduce additional decoherence channels due to voltage fluctuations~\cite{friesen:03,hu:06} or sample heating. Whether this will affect the control performance will have to be determined experimentally.

In summary, we have employed optimal quantum control theory to design quantum gates for solid-state qubits interacting with their environment. For a gate-controlled semiconductor double quantum dot subject to phonon couplings, we have shown that optimized gates can strongly suppress phonon-assisted decoherence and can boost the fidelity by several orders of magnitude. We attribute our finding to the fact that in the process of decoherence it takes some time for the system to become entangled with its environment. If during this entanglement buildup the system is acted upon by an appropriately designed control, it becomes possible to channel back quantum coherence from the environment to the system. We therefore believe that our findings are relevant for a much broader class of solid state systems where quantum information is encoded in long-lived quasi-groundstates with small energy separations, such as electon or nuclear spins, resulting in slow scattering processes that can be manipulated by means of quantum control.

Work supported in part by the Austrian Science Fund FWF under projet P18136--N13.


\begin{thebibliography}{27}
\bibitem[{rem()}]{remark.densitymatrix}
\bibinfo{note}{Within our density-matrix framework we keep as dynamic variables the Bloch vector $\langle\bm\sigma\rangle$, with $\bm\sigma$ the vector of Pauli matrices $\sigma_1=\begin{pmatrix} 0&1\\1&0\\ \end{pmatrix}$, $\sigma_2=\begin{pmatrix} 0&-i\\i&0\\ \end{pmatrix}$, and $\sigma_3=\begin{pmatrix} 1&0\\0&-1\\ \end{pmatrix}$, the coherent phonon amplitude $\langle b_{\bm q}\rangle$, the phonon-assisted density matrix $\langle\!\langle \bm\sigma b_{\bm q}\rangle\!\rangle$, and the phonon occupation $\langle\!\langle b_{\bm q}^\dagger b_{\bm q}^{\phantom\dagger}\rangle\!\rangle$. Other cumulants, such as two-phonon correlations $\langle\!\langle b_{\bm q'} b_{\bm q}\rangle\!\rangle$, turned out to be of only minor importance, and were therefore neglected. The time evolution of the dynamic variables is obtained by direct integration of the equations of motion for the cumulants. See refs.~\cite{rossi:02,hohenester.prl:04,hohenester.review:06} for further details.}
\end{thebibliography}
\end{document}